%% file: preprint.tex
\documentstyle[pre,aps]{revtex}
\input{epsf}

\begin{document}

\title      { 	Modeling noisy time series: \\ Physiological tremor }
\author {       J. Timmer   \\ 
		Fakult\"at f\"ur Physik, Hermann-Herder Str.~3 \\
		79104 Freiburg, Germany  \\  
                Zentrum f\"ur Datenanalyse und Modellbildung,\\
                Eckerstr.~1, 79104 Freiburg, Germany \\ 
		e-mail: jeti @ fdm.uni-freiburg.de
                         }

\maketitle

%PACS numbers: 5.45+b, 2.50.Ey, 87.10+e

%%\clearpage

\begin {abstract}

Empirical time series often contain observational noise. We investigate
the effect of this noise on the estimated parameters of models fitted 
to the data. For
 data of physiological tremor, i.e.~a small amplitude oscillation of the
outstretched hand of healthy subjects, we compare the results for a
linear  model that
explicitly includes additional observational noise to one that ignores this noise.
We discuss problems and possible solutions for nonlinear deterministic
as well as nonlinear stochastic processes.
Especially we discuss the state space model applicable for modeling
noisy stochastic systems and Bock's algorithm capable for modeling
noisy deterministic systems.

\end {abstract}
%\twocolumn
%%\clearpage
\draft
\section { Introduction }
One of the aims in time series analysis of complex systems is to 
find a dynamical model for the data \cite{crutchfield87}.
A well fitting model might yield insight into the underlying process.
But also if prediction is the main goal, e.g.~in order to discriminate chaos
from stochasticity \cite{sugihara90,casdagli91}, or if fitted models are
used to determine dynamical invariants of the system 
\cite{kadtke93,aguirre95}, an optimal estimation of the parameters 
is desirable. 
Our own interest was stimulated by modeling physiological time series
in order to gain insight into the underlying systems.

The dynamical models can be (time-continuous) differential or
(time-discrete) difference equations. Methods to estimate
parameters in differential equations can be subdivided 
according to whether they require an estimation of the derivatives
from the data \cite{cremers87,gouesbet91} or not
\cite{bock83,edsberg95}. Difference
equation allow to use a great variety of different approaches  
to model the mapping from past values to the present one. These range
from parametric ones \cite{giona91,aguirre95} via neural 
nets  \cite{lapedes87,weigend90,principe92}, radial basis 
functions \cite{casdagli89,poggio90} and nonparametric 
models \cite{tjoestheim94} to local linear models \cite{casdagli91}. 

For all these methods a significant amount of observational noise
 can be a severe problem.
Especially for difference equations the functional relation between
 past and present values will
be underestimated if observational noise is not included in the model.
In the first part of this paper
we exemplify this for data of physiological human hand tremor. These
data contain up to 50 \% observational noise.
Inspired by the physics of the process, linear stochastic
 autoregressive (AR) processes 
have already been suggested in 1973 to model these data \cite{randall73}. 
In \cite{gantert92} these data were analyzed by a linear state space
model which takes the observational noise into account.
For one data set we compare these two approaches in detail.
In the second part of this paper we 
discuss problems introduced by observational noise and possible
 solutions for non-linear deterministic
as well as non-linear stochastic processes.

\section {  Modeling Physiological Tremor } \label{model_phys}

The outstretched hand of a healthy subject exhibits a small amplitude
oscillation called physiologic tremor \cite{deuschl96}. The movement 
of the hand can be measured by piezoresistive accelerometers attached at
the hand. Simultaneously the flexor and extensor muscle activity is
recorded. The spectra of the muscle activity data are often flat in
physiological tremor reflecting an uncorrelated activity of
motoneurons \cite{timmer98a}. 
For an analysis of physiological tremor data in which 
 the muscle activity is correlated and included in the
modeling, see \cite{timmer98b}.

\begin{center}
\begin{figure} [t]
  \input {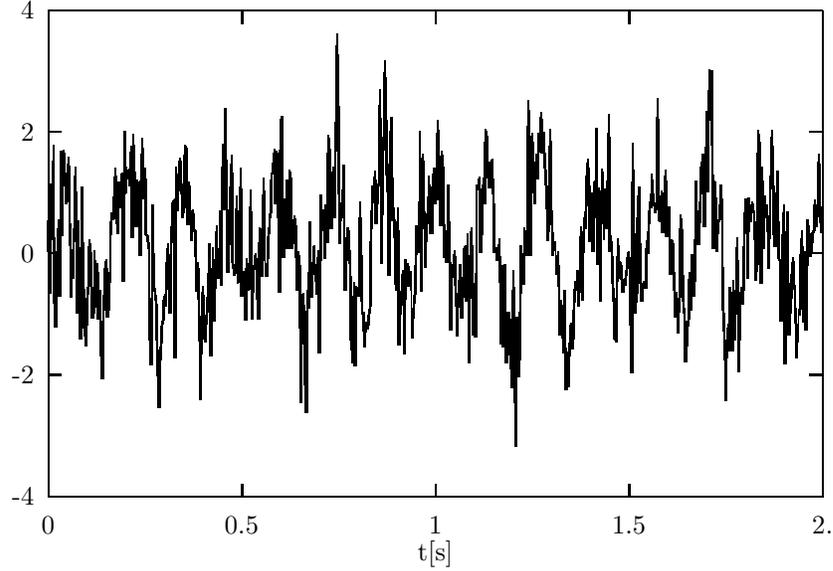}
    \caption{\label{data_fig} A 2 s section of human physiological
	hand tremor data. }
\end{figure}
\end{center}

\begin{center}
\begin{figure}
  \input {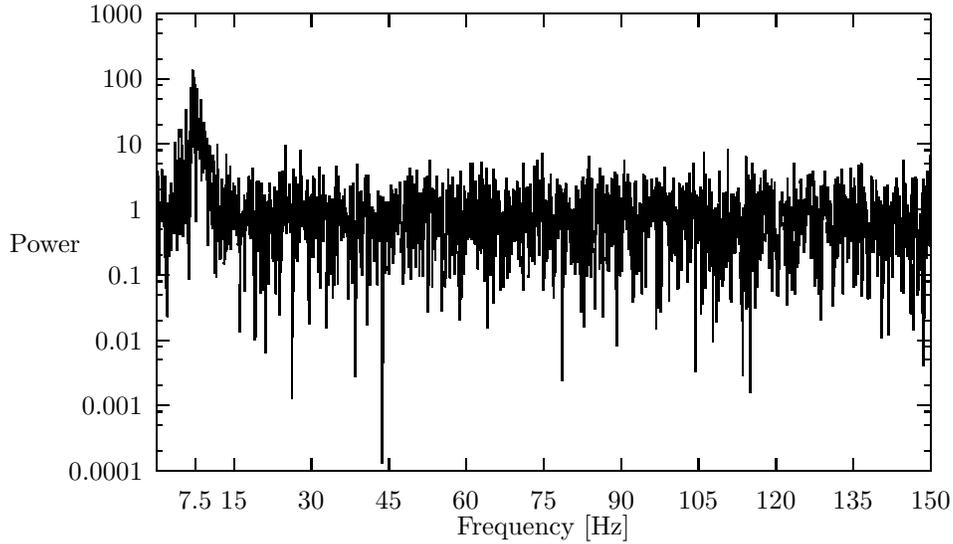}
    \caption{\label{perio_fig} Periodogram of the physiological tremor data set. }
\end{figure}
\end{center}

Fig.~\ref{data_fig} displays a 2 s section of the whole
35 s measurement which is sampled with 300 Hz. 
%
%\marginpar {Please Fig.~\ref{data_fig} around here}
%------------------------------------------------------------
%  Fig. 1 (FIG1.tex) around here
%------------------------------------------------------------
%
The time series consist of 10240 data points
covering approximately 250 periods of the oscillation. The data are
normalized to zero mean  and unit variance.
Fig.~\ref{perio_fig} displays the periodogram, i.e.~the absolute 
%
%\marginpar {Please Fig.~\ref{perio_fig} around here}
%------------------------------------------------------------
%  Fig. 2 (FIG2.tex) around here
%------------------------------------------------------------
%
value of the  Fourier transform squared. For the treatment of 
special problems of estimating the spectrum from the periodogram 
for tremor time series, see \cite{timmer96}. The periodogram shows 
on the one hand the high amount of observational noise in these
data. Roughly estimating the variance of the signal by summing up the
periodogram in the range from 2.5 to 12.5 Hz, and that of the noise 
from the remaining frequencies results in a signal-to-noise ratio
of 0.93 in relative amplitudes.
On the other hand the periodogram shows one broad peak around a frequency
of 7.5 Hz. This peak is usually explained as a resonance 
phenomenon \cite{stiles80}. 
The outstretched hand is a damped oscillator which is excited 
by the uncorrelated muscle activity. Usually no higher harmonics show 
up in the periodogram indicating that the process is linear. 

The autoregressive (AR) processes invented
by Yule \cite{yule27} are expected to fit such data well.
 An AR process of order $p$ is given by :
\begin{equation} 
  x(t) = \sum_{i=1}^{p} a_{i} \, x(t-i) \, + \, \epsilon(t) \quad ,
\end{equation}
where $ \epsilon(t) $ denotes a uncorrelated Gaussian distributed
random variable with mean zero and variance
$\sigma^2$.  Such a process can be interpreted,
depending on the chosen parameters, as a combination of relaxators and
damped oscillators \cite{honerkamp_buch}.
For example, an AR process of order 2 that corresponds to a damped 
oscillator with
characteristic period $T$ and relaxation time $\tau$ given by:
\begin{eqnarray}
  a_1 & = & 2 \cos \left(\frac{2\pi}{T}\right) \, \exp \,(-1/\tau)\label{a1}\\ 
  a_2 & = & - \exp \,(-2/\tau)\label{a2}  \qquad .
\end{eqnarray}
The spectrum is given by :
\begin{equation}
S(\omega) = \frac{\sigma^2}
	{| 1 - a_1 e^{-i\omega} - a_2 e^{-2i\omega} |^2} \qquad .
\end{equation}

AR processes can be generalized to the autoregressive moving average
(ARMA[p,q]) processes by including past q driving noise terms in the
dynamics. For theoretical reasons ARMA[p,p-1] should be preferred to
AR[p] processes for modeling of sampled continuous-time processes 
\cite{phadke74}. Experience shows that differences in the results are small.

A more substantial generalization is the linear state space model  (LSSM)
\cite{honerkamp_buch} which enables the explicit modeling of observational 
noise $\eta(t)$ that contributes to the measured $y(t)$ :
\begin{eqnarray}
  \vec{x}(t) & = & {\bf{A}} \, \vec{x}(t-1) + \vec{\epsilon}(t), \qquad 
	\vec{\epsilon}(t) \in {\cal{N}}(0,{\bf{Q}})   \label{zrm1}\\ 
  y(t) & = & {\bf{C}} \, \vec{x}(t) + \eta(t),  \qquad 
	 \eta(t) \in {\cal{N}}(0,R) \label{zrm2}  \qquad .
\end{eqnarray}
Eq.~(\ref{zrm1}) describes the linear dynamics. Eq.~(\ref{zrm2}) maps the
dynamics to the observation including the observational noise $\eta(t) $.
Another advantage of the LSSM is its capability to model superpositions
of linear processes. This is not possible for AR or ARMA processes.
 The spectrum of a LSSM is given by :
\begin{eqnarray}
      S(\omega) = {\bf{C}} ({\bf{1}}-{\bf{A}}e^{-{\rm i} \omega})^{-1} {\bf{Q}}
      \left(({\bf{1}}-{\bf{A}}e^{{\rm i} \omega})^{-1}\right)^{\rm T} 
                                                      {\bf{C}}^{\rm T} + R
    \qquad .   \label{eq:arsp}
\end{eqnarray}
The superscript $T$ denotes transposition. Spectra of AR or ARMA processes
are special cases of Eq.~(\ref{eq:arsp}).

While parameter estimation in AR models e.g.~by the Burg or
Durbin-Levinson algorithm
is well established, parameter estimation in the LSSM is more
cumbersome, see \cite{honerkamp_buch} for a detailed description.
Usually the Expectation -- Maximization (EM) algorithm is applied 
\cite{dempster77,shumway82}. 
The EM - algorithm is a general iterative procedure to estimate parameters
for models in which not all variables are observable, here $ \vec{x}(t)$.
Denoting the joint density of $ \vec{x}(t)$ and $ y(t)$
given the parameters $\Theta$ by $p(x,y|\Theta)$ and  the density of
 $ \vec{x}(t)$ given the data  $ y(t)$ and the  parameters 
$\Theta$ by $p(x|y,\Theta)$ the quantity :
\begin{equation}
  L(\Theta,\Theta^i) = < \mbox{ln} (p(x,y|\Theta) )>_{p(x|y,\Theta^i)}
\end{equation}
is calculated in the $i$th Expectation step.
In the Maximization step, $ L(\Theta,\Theta^i) $ is maximized with
respect to $ \Theta $ yielding  $ \Theta^{i+1} $.

In the case of the LSSM, this means that starting from some initial
 values for the parameters ${\bf A}, {\bf Q}, {\bf C}, R$ the hidden dynamic
 variable $ \vec{x}(t)$ is
estimated by the Kalman filter \cite{kalman60} in the expectation step. 
Denoting the estimator of a quantity $z(t)$ based on the data
 $y(1),...,y(t')$ by $z_{t|t'} $, the covariance matrix of the estimated
 $ \vec{x}(t)$ by $\Omega_{t|t'}$ and the variance of the prediction
 errors $(y(t)-y_{t|t'})$ by $\Delta_{t|t'} $ the Kalman Filter reads:
\begin{eqnarray}
\Omega_{t|t-1} & = & {\bf A} \Omega_{t-1|t-1} {\bf A}^T +  {\bf Q}  
						\label{kalman1} \\
\Delta_{t|t-1}   & = &  {\bf C} \Omega_{t|t-1}  {\bf C}^T +R \\
K   & = & \Omega_{t|t-1}  {\bf C}^T \Delta_{t|t-1}^{-1} \\
\Omega_{t|t}  & = &  (1- K  {\bf C}) \Omega_{t|t-1}\\
 \vec{x}_{t|t-1} & = & {\bf A} \vec{x}_{t-1|t-1} \\
y_{t|t-1} & = &  {\bf C} \vec{x}_{t|t-1} \\
 \vec{x}_{t|t} & = &  \vec{x}_{t|t-1} + K (y(t)-y_{t|t-1}) \label{kalman2}
\end{eqnarray}
Since the model is gaussian and linear, the density  $p(x|y,\Theta)$
is completely specified by $ \vec{x}_{t|t} $ and their covariances
$ \Omega_{t|t}$. Note, that the covariances $ \Omega_{t|t}$ only
depend on the model parameters, not on the data. 
An improvement of the estimated quantities by the
so-called smoothing filter and the lengthly equations for the
parameter update of  $ {\bf A}, {\bf Q}, {\bf C}, R$ in the
Maximization step are given in \cite{honerkamp_buch}. This 
procedure is iterated until some convergence criterion is fulfilled.

There are different criteria to judge the goodness of fit of a
dynamical model.
\begin{itemize}
\item Whiteness of the prediction errors. 
	The model should explain all correlations in the data yielding
	uncorrelated prediction errors.
	By the Kolmogorov - Smirnov test \cite{recipes}
	it can be tested whether
	the periodogram of the prediction residuals is consistent with white 
	noise \cite{brockwell87}.
\item A knee in the variance of the prediction errors for ascending  
	model order indicates the correct order. Criteria like 
	AIC \cite{akaike73},
	BIC \cite{schwartz78}, MDL \cite{rissanen78} can be used to take into 
	account the different
	number of parameters in the compared models. The application of these
	criteria is not without problems \cite{aguirre94}. 
	If the model class is not correct,
	these criteria will choose "some" order. 
\item   The distribution of some feature can be derived from 
	numerous realizations
	of the model and the compatibility of the value of
	 the feature calculated from the
	data with this distribution can be examined, see e.g.~\cite{witt94}.
\end{itemize}

In the case of linear modeling there are two more criteria.
\begin{itemize}
\item Since the parameters in linear models are related to relaxation
	times of the corresponding oscillators and relaxators,
	negligible relaxation times indicate a to large model order.
\item According to Eq.~(\ref{eq:arsp}) the spectra of linear processes
	can be calculated from the parameters of the estimated models.
	Since the spectrum is the expectation of the periodogram,
	a comparison of the spectrum to the periodogram of the data 
	 serves as a qualitative criterion. 
\end{itemize}

We now present the results of the analysis of the data shown in 
Fig.~\ref{data_fig}. We do not give the estimated parameter here since they are
rather uninformative. Instead, 
Tab.~1
%Tab.~\ref{tsundtaus} 
displays the
resulting periods and relaxation times. These characteristic times can
be calculated for any model order by formulas analogous to 
Eqs.~(\ref{a1},\ref{a2}) \cite{honerkamp_buch}.
%

%\begin {table} [t]
\begin{center}
\begin {tabular}{|c|c|c|c|c|c|c|}
\hline
    model order   & \multicolumn{3}{c|}{AR model} &  
   		\multicolumn{3}{c|}{LSS model}  \\
\hline
       & relaxation &  period   & prob. & relaxation &  period  & prob.     \\
       &  time      &           &  WN  & time        &         &  WN  \\
\hline
    1   &    1.2      &   -  &     0    & 7.9   &  -  &         0  \\
\hline
    2   &    2.9      &   -      & 0    & 91.9    &   39.9 & 0.24 \\
        &    1.0     &     -    &            &    -      &     -  &   \\
\hline
   3    &     4.8    &    -   &   10$^{-8}$  &  91.9    &  39.9 & 0.27  \\
       &      1.3   &  6.5      &           &  0.2     &   -  &       \\
\hline
  4     &   6.6      &   -      &  10$^{-7}$ & 91.9       &   39.9 & 0.23 \\
       &    1.6     &    4.3     &            & 9.0     &   8.3    &    \\
       &    1.4     &      -    &              &  -       &  -      & \\
\hline  
 5       &  7.5       &   -      & 10$^{-8}$  & 91.9     & 39.9  & 0.24  \\
	&   1.5      &   4.3      &           &  9.4      &  8.4  &     \\
       &    1.2     &   11.9      &            &   0.6     &   -  &      \\
\hline
\end {tabular}
\vskip 0.2cm
%\caption
{  TAB. 1. Relaxation times and periods of the fitted
	AR and LSS models of ascending order for the physiological 
	tremor data. "prob. WN" denotes the probability of the
	residuals to be consistent with white noise.   }
%\label{tsundtaus}
\end{center}
%\end {table}

For an estimated oscillator a relaxation time smaller than
%that is in the order of magnitude of 
the period of the oscillation indicates that this
oscillator is insignificant. Pure relaxators with an relaxation time of 
a few time steps have also to be regarded as negligible. 
For the LSSM, a model order of two with a period of approximately 40
time steps, corresponding to a frequency of 7.5 Hz, 
and a relaxation time of 91 time steps is clearly identified.
For the AR models no significant structure is detected.
%
%Tab.~\ref{tsundtaus} 
Tab.~1
also gives the probability of the residuals
to be consistent with white noise. For the LSSM, models with an order
larger than one yield prediction residuals which are consistent 
with white noise. The residuals of the AR processes are not consistent
with white noise for all model orders investigated.

Fig.~\ref{vari_fig} displays the variance
%
%\marginpar {Please Fig.~\ref{vari_fig} around here}
%------------------------------------------------------------
%  Fig.3 (FIG3.tex) around here
%------------------------------------------------------------
%
of the prediction error in dependence on the model order.
The graph for the LSSM exhibits a clear cut knee at an order of 2
whereas the graph of the AR model does not allow for a clear cut
decision. The graph saturates around order 5 and decreases slowly
for larger orders. We fitted AR model up to order 50 and applied the
AIC. Even for this unrealistic high orders the function AIC(p)
still decreases.

\begin{figure}
\begin{center}
\hspace{1.5cm}\epsfxsize=0.5\textwidth \epsfbox{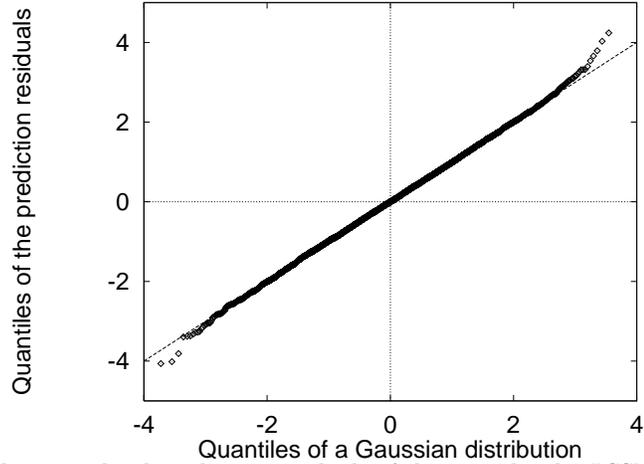}
    \caption{\label{qq_fig}Quantile-quantile plot of the normalized 
	prediction residuals of the second order LSSM with
	respect to a Gaussian distribution. The straight gives the
	expected behavior in the case of Gaussianity.}
\end{center}
\end{figure}

Fig.~\ref{qq_fig} shows a quantile-quantile --  plot with respect to a Gaussian
distribution of the normalized prediction residuals for the LSSM
%
%\marginpar {Please Fig.~\ref{qq_fig} around here}
%------------------------------------------------------------
%  Fig.4 (QQ1.eps) around here
%------------------------------------------------------------
%
of order two and the expected straight line. Only ten of the 10240 
points deviate from the straight line indicating Gaussianty of the
prediction residuals.

\begin{center}
\begin{figure}
  \input {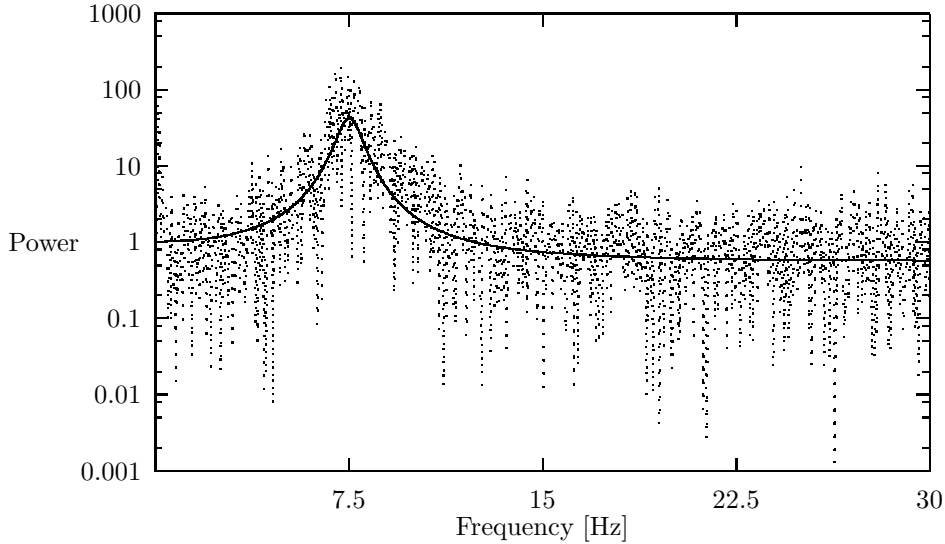}
    \caption{\label{spec_per_fig}Periodogram of the data (dotted) and spectrum of
	the fitted second order linear state space model (solid).}
\end{figure}
\end{center}

All these results indicate that the LSSM of order two describes the
data adequately. Fig.~\ref{spec_per_fig} displays the
%
%\marginpar {Please Fig.~\ref{spec_per_fig} around here}
%------------------------------------------------------------
%  Fig.5 (FIG5.tex) around here
%------------------------------------------------------------
%
low frequency part of
spectrum estimated from the parameters of the fitted LSSM of order two
and the periodogram of the data. This confirms the above results.

\begin{center}
\begin{figure}
  \input {FIG3.tex  }
    \caption{\label{vari_fig} Residual variance of the fitted AR (+)
	and LSS ($\Diamond$) models in dependence on the model order.}
\end{figure}
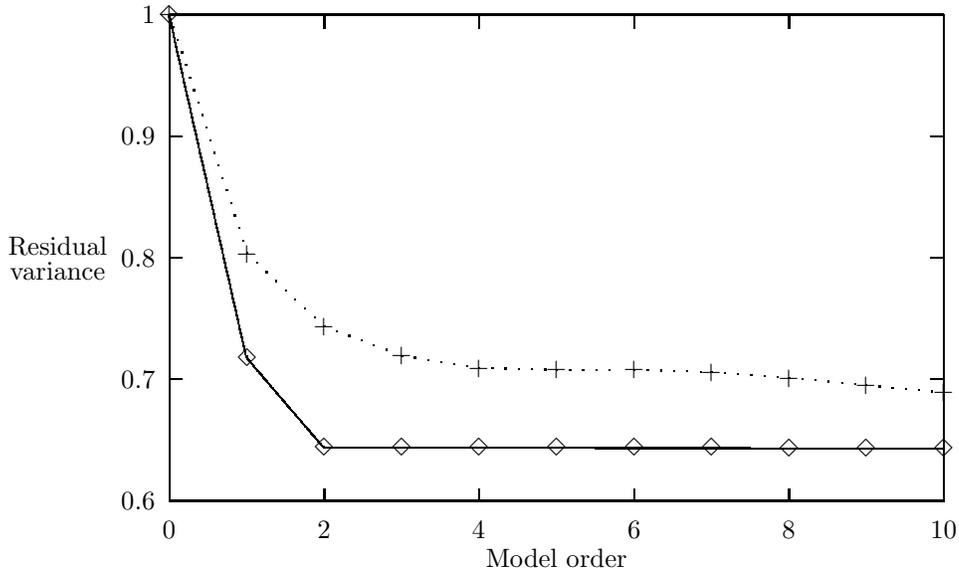
\end{center}

In terms of physics and physiology these result show that the
physiological tremor is a linear damped oscillation driven by 
uncorrelated muscle activity. The parameters of the model
describe the mechanical properties of the muscle-hand system,
i.e. its mass, stiffness and damping.
\section {  Modeling Noisy Time Series   } \label {modeling}
The behavior of the AR model in the previous section, i.e.~not
recovering the correct model order, is caused by the
observational noise. Whereas the LSSM includes the observational noise
explicitly in the model, the AR model assumes the data to be free from 
observational noise.
We use a simple first order process to demonstrate the consequences. 
In an AR[1] model
\begin{equation}
  x(t)=a x(t-1) + \epsilon(t)
\end{equation}
the parameter $a$ can be estimated without bias by :
\begin{equation} \label {ar1_schaetz}
  \hat{a} = \frac{ \sum x(t-1) x(t)} { \sum x(t-1) x(t-1) } \quad .
\end{equation}

If the dynamics is covered by observational noise:
\begin{equation}
  y(t)=x(t) + \eta(t), \qquad \eta(t) \sim {\cal{N}}(0,R) \quad ,
\end{equation}
the expected value of $ \hat{a} $ estimated  analogously to
 Eq.~(\ref{ar1_schaetz}) from $y(t) $ is
\begin{equation} \label {ar1_unterschaetz}
  <\hat{a}> = \frac {< y(t-1) y(t)> }{ <y(t-1) y(t-1)> } 
	 = a \,	\frac{1}{ 1 + \frac{ R} {< x(t)^2>}}		\quad .
\end{equation}
Thus, the parameter $a$ is underestimated. The degree of the 
underestimation depends on the signal-to-noise ratio.
This effect is known from linear regression theory where it
is called the "Error-in-variables" - problem \cite{fuller87} and was
first mentioned in the context of time series in \cite{kostelich92}.
For the data analyzed in the previous section the variance of the
signal nearly equals the variance of the noise. Thus, an
underestimation of $a$ by a factor of two is expected. Indeed,
even for the misspecified first order models,
the parameter estimated from the LSSM is $0.882$ and that for the
AR[1] model is $0.443$. Note, that in the case of second order models
the misestimation of the parameters for the AR[2] models leads
to a detection of two relaxators, whereas the LSSM recognizes the
oscillator, see Tab.~1. Thus the AR model leads to a wrong
interpretation.

The underestimation of the functional relation between past and
present values in the case that observational noise is not 
taken into account carries over to more general models. 
Assuming stationarity of the underlying process systems can
be linear or nonlinear, resp.~deterministic or stochastic.
Our approach to discriminate deterministic from stochastic
systems, i.e.~the presence of dynamical noise, is rather practical.
For example, in the data analyzed in the previous section
it is imaginable to model the muscle activity by a deterministic
system, including a model for the brain and the spinal tract.
This would result in a very high dimensional dynamical system
whose output is equally well captured by white noise. Thus,
whenever a huge amount of influences enters the system
significantly a stochastic description might be recommended.
But also if there exists no shadowing trajectory for a realization of 
deterministic
system, the use of stochastic models might be indicated \cite{dawson94}.

The dynamics can be described by differential equation or
by maps. In the following we discuss possible treatments
of observational noise in the different cases. If present
the dynamical noise is assumed to be Gaussian and  additive.
By the latter assumption we exclude models which are able to
describe fluctuations in the parameters, e.g.~bilinear 
models \cite{subbarao84}.
\subsection{ Linear deterministic case}
The linear deterministic case is easy to treat since the
solutions of the dynamical equations are explicitly known
and can be fitted to the noisy data. 
Parameters of continuous- and discrete-time
 modeling are connected by equations like Eq.~(\ref{a1},\ref{a2}). 
An application to physiological data is given in \cite{groothuis91}.
\subsection{ Linear stochastic case}
The linear stochastic case is solved by the state space model
discussed above.
Because of the linearity and gaussianity of the model there
is a direct connection between the parameters from discrete-time and
those from the continuous-time models. Parameter estimation for the
continuous-time case is presented in \cite{singer93}.
An application of the state space model to astrophysical 
data is given in \cite{koenig97}.

\subsection{ Nonlinear deterministic case} \label {ndc}
 The effect of observational
noise is different for fitting differential and difference equations
to the data.

For modeling data by differential equations two approaches can
be distinguished depending on whether time derivatives of the process has
be estimated from the data or not. Since estimating derivatives
form the data amplifies the noise, the former method as applied in
\cite{cremers87,breeden90,eisenhammer91,gouesbet91,gouesbet92,irving97}
is vulnerable to significant amounts of 
observational noise as demonstrated in \cite {irving97}.

To fit differential equations to noisy deterministic
data without estimating derivatives from the data,
 at least two different approaches exist. In the initial value approach
one chooses some initial parameters in the model and an initial value
for the trajectory, integrates the differential equation and tries
to minimize the prediction error by some minimization 
algorithm \cite {edsberg95}.
Without many precautions, this procedure has been shown to be unstable, 
i.e.~yielding divergent trajectories, and is susceptible to
stopping in a local minimum \cite{richter92,timmer98c}. A more 
sophisticated algorithm was developed by Bock \cite{bock81,bock83}.
His elegant multiple shooting approach starts with a discontinuous
trajectory which stays close to the data. The continuity of the
underlying trajectory enters into the algorithm by a constraint in the
cost function. This constraint is nonlinear in the parameters but
enters the optimizing strategy only in a linearized way. Therefore,
the trajectory is allowed to be discontinuous at the beginning of
the iteration but is forced to become continuous in the end.

\begin{figure}
\begin{center}
\parbox{\textwidth}{
\epsfxsize=0.4\textwidth \epsfbox{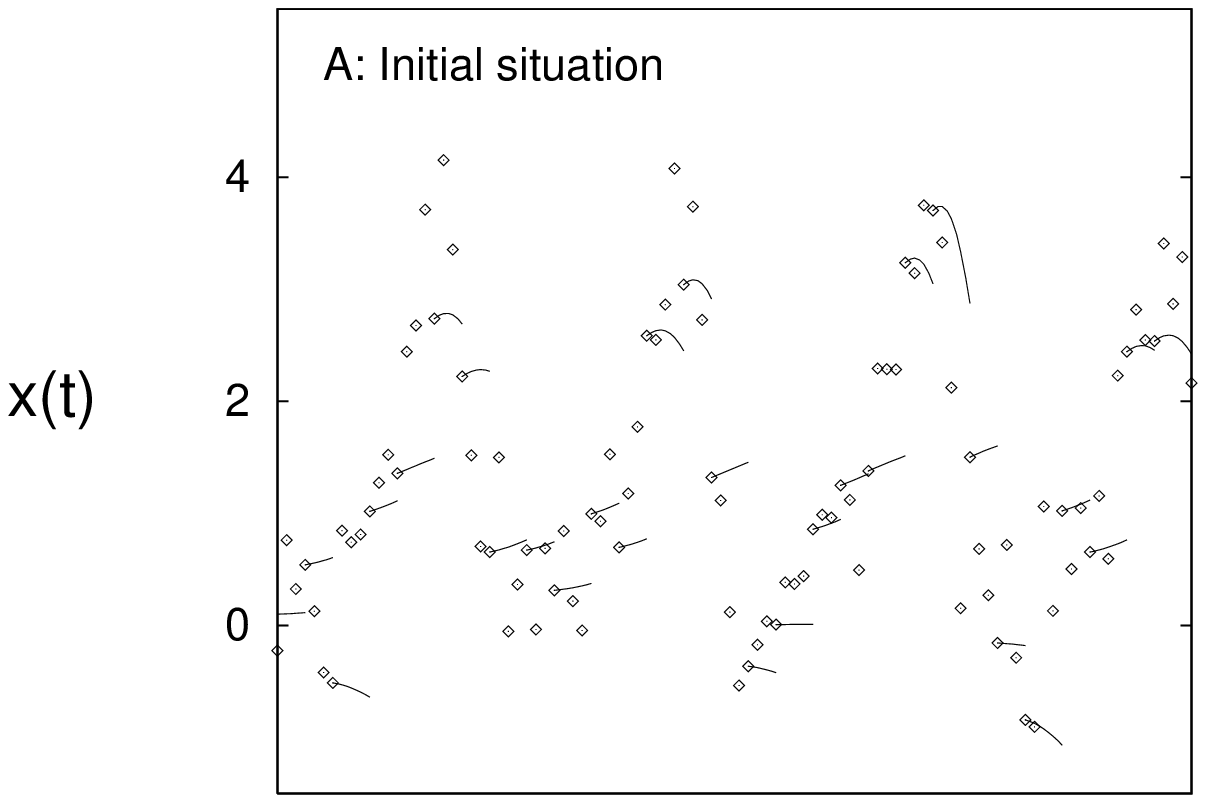}}
\parbox{\textwidth}{
\epsfxsize=0.4\textwidth \epsfbox{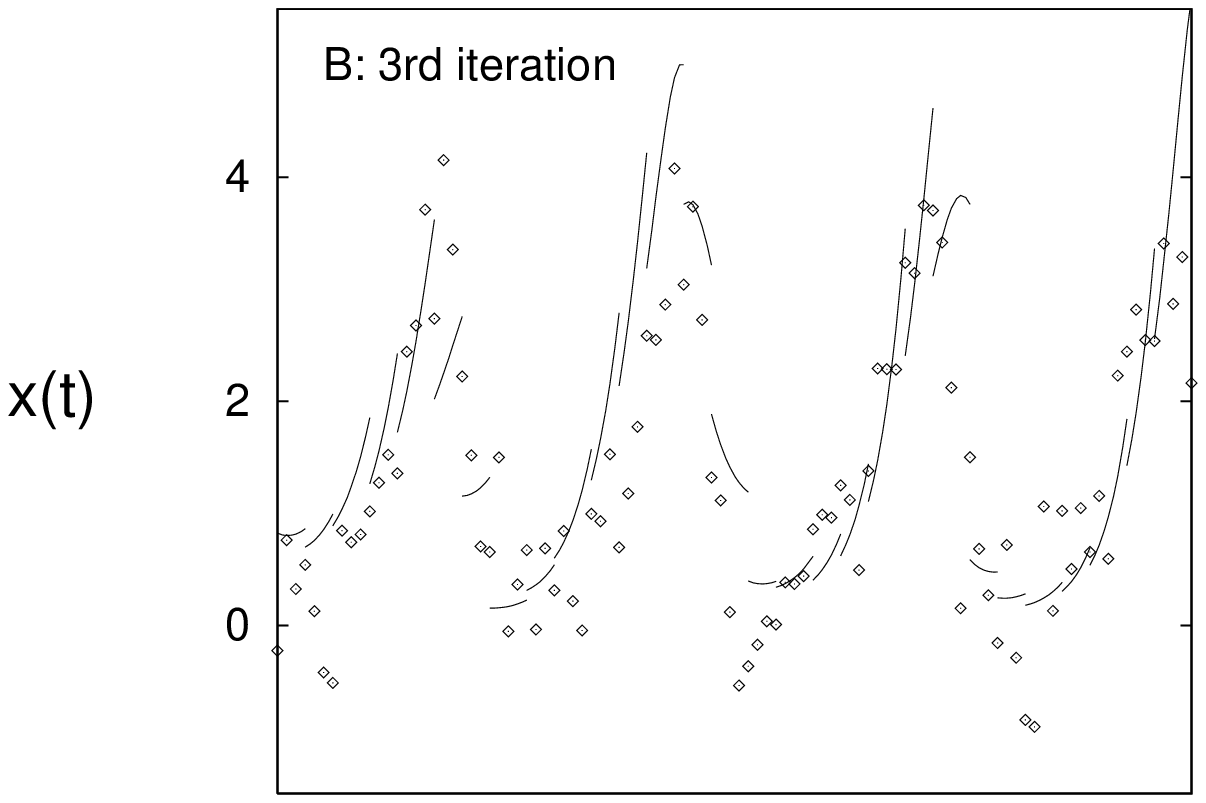}}
\parbox{\textwidth}{
\epsfxsize=0.4\textwidth \epsfbox{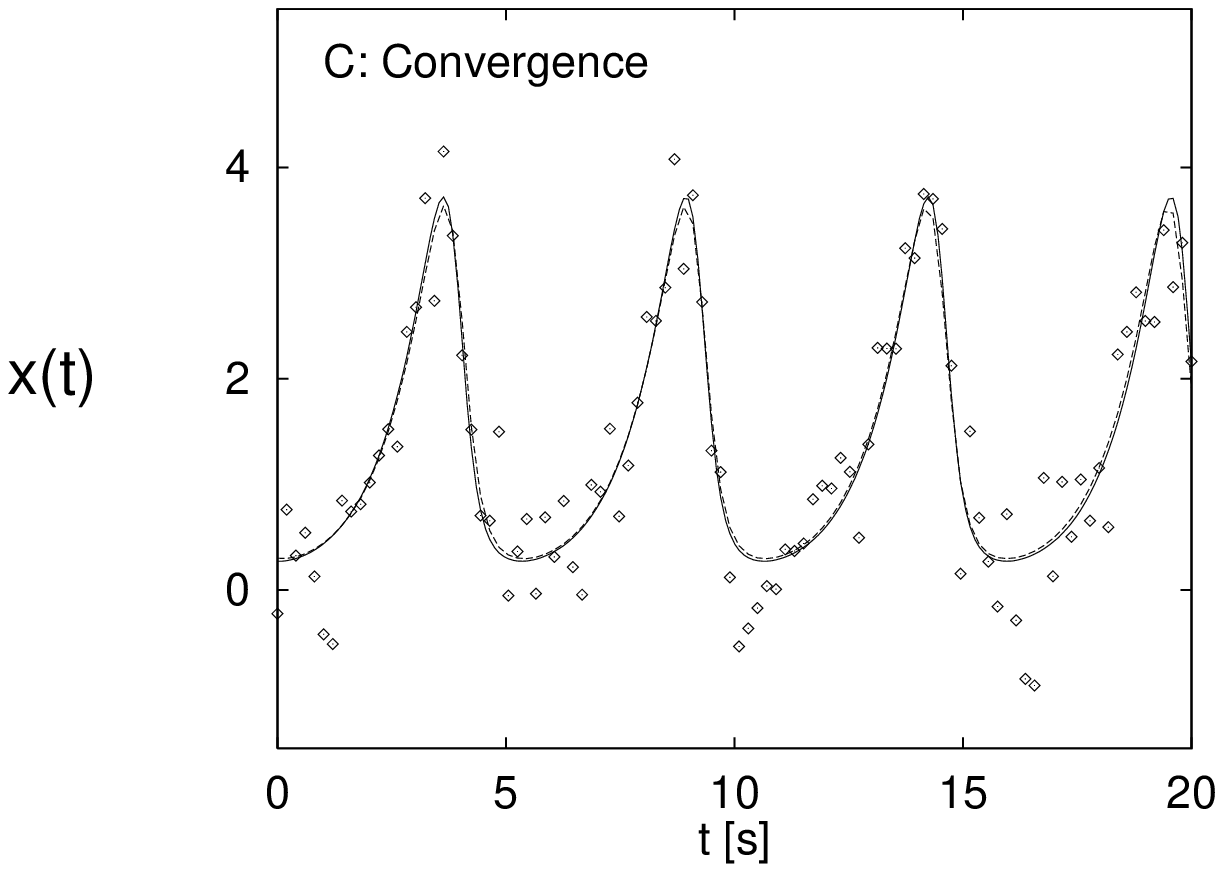}}
 \caption{\label{bock_fig}Convergence of Bock's algorithm for the noisy
	Lotka-Volterra system. A: Initial situation. B: After the
	third iteration. C: After convergence, dashed line: true
	trajectory.}
\end{center}
\end{figure}

We illustrate the behavior of Bock's algorithm by the restricted 
Lotka-Volterra system which has been used to compare different fitting
 procedures \cite{bockdiss,varah82,tjoa91,edsberg95}:
\begin{eqnarray}
\dot{x}_1 & = & k_1 x_1 - k_2 x_1 x_2 \\
\dot{x}_2 & = & k_2 x_1 x_2 - k_3 x_2
\end {eqnarray}
with the parameters $k_1= 1.0$, $k_2= 1.5$ and $k_3= 2.0$ and initial 
values $x_1(0)= 0.3$, $ x_2(0) = 0.8$. The system is integrated by the
routine BSSTEP from \cite{recipes} for 20 s and sampled with $\Delta t
= 0.2$ s. The standard deviation of the data is 1.09.
Observational noise with standard deviation of 0.5 was added to the
signal. To fit the parameters, only the first component
was used. All starting
points for the second component were set to 0.5. Initial values 
for the parameter estimates were $\hat{k}_1= 0.5$, $\hat{k}_2=
0.75$ and $\hat{k}_3= 1.0$. 30 starting points were used for the
initially discontinuous trial trajectory.
 Fig.~\ref{bock_fig} shows initial situation (A), the third iteration
(B) and the converged solution (C) after 16 iterations  of the algorithm.
The true trajectory is reproduced well.
%
%\marginpar {Please Fig.~\ref{bock_fig} around here}
%------------------------------------------------------------
%  Fig.6 (LOVO1.eps,LOVO2.eps2,LOVO3.eps) around here
%------------------------------------------------------------
%
With respect to the confidence regions, the estimated parameters
are compatible with the true parameters: $\hat{k}_1= 1.01 \pm 0.14$,
 $\hat{k}_2=1.53 \pm 0.34$ and $\hat{k}_3= 2.02\pm 0.45$.

Simulation studies comparing both algorithms show that Bock's 
algorithm is more stable and converges to the global minimum for a
larger set of initial guesses for the parameters than the
initial value approach \cite {richter92,timmer98c}. 
Note, that for both approaches it is not necessary to reconstruct
the dynamics by delay-embedding \cite{takens81} avoiding all problems
related to choosing the delay time \cite {kugiumtzis96}. 

In a simulation study Bock's algorithm has been successfully
applied to model noisy chaotic data from dissipative as well as
conservative systems \cite{baake92a,kallrath}. Applications of
this method to chemical and physiological data are reported 
in \cite{bock81,baake92b}. The basic idea of Bock's algorithm,
i.e.~taking advantage of the fact that the process under investigation
has produced a continuous trajectory, carries over to modeling
noisy deterministic systems by maps.
Nevertheless this idea has not attracted much attention and 
modeling data by maps is generally treated as a regression
not as a dynamical problem.

Stimulated by \cite{crutchfield87} a large number of
different approaches have been suggested to model
 nonlinear dynamics by maps. They are distinguished by the
type of basis function used to approximate the nonlinear functional 
relationship between past values and the present one. 
Amongst others, polynomials \cite{giona91}, 
sigmoidals \cite{lapedes87,weigend90,principe92},
 radial basis functions \cite{poggio90} and local linear models
\cite{casdagli91} have been suggested.
Analogous to the AR model discussed above all these methods
are subject to the ''Error-in-variables'' -- problem.
For the tremor data and the AR model of order two in
Sec.~\ref{model_phys}  above this problem led to 
a detection of two relaxators instead of an oscillator. In
\cite{kadtke93} it was reported that observational noise covering
chaotic data led to a detection of a periodic process. That might 
be caused by the same phenomenon of underestimated parameters.

The generalization of the theory of ''Error-in-variables'' from nonlinear
regression, see \cite{carroll95} for a recent review, to modeling of noisy
nonlinear deterministic time series by maps
 and an application to empirical data are presented in \cite{kantz96}.

%\centering

\subsection{ Nonlinear stochastic case}
For the nonlinear stochastic case, including a possible nonlinear
mapping $ g(.) $ of the state vector $ \vec{x}(t)$ to the 
observation $ y(t)  $ :
\begin{eqnarray}
  \dot{\vec{x}}(t) & = &  \vec{f}(\vec{x}(t)) + \vec{\epsilon}(t), \qquad 
	\vec{\epsilon}(t) \in {\cal{N}}(0,{\bf{Q}})   \label{nlzrm1}\\ 
  y(t) & = & g({\vec{x}}(t)) \,  + \eta(t),  \qquad 
	 \eta(t) \in {\cal{N}}(0,R) \label{nlzrm2}  \qquad ,
\end{eqnarray}
a generalization of the Kalman filter
Eq.~(\ref{kalman1}-\ref{kalman2}) appears to be the solution.
Therefore, for the discrete-time version, the
matrices $\bf {A} $ and $\bf{C} $ are replaced by linear
approximations of  $\vec{f}(\vec{x}(t))$ and $g({\vec{x}}(t))$ 
\cite{gelb89,mendel95}:
\begin{eqnarray}
\bf {A} & = & \frac{\partial \vec{f}} { \partial \vec{x}} 
			\left|_{\vec{x}=\vec{x}_{t-1|t-1}} \right. \\
\bf{C} & = & \frac{\partial g } { \partial \vec{x}} 
			\left|_{ \vec{x}=\vec{x}_{t|t-1}}  \right.
\end{eqnarray}
This solution is, in general, not valid for two reasons. 
First, as mentioned in Sec.~\ref{model_phys},
the variance $\Omega_{t|t}$ of the predicted state variable 
$\vec{x}_{t|t}$ depends only on the model, i.e.~not on the amount of
data, and might be so large that the linearization is not valid.
Thus, the distribution of $\vec{x}_{t|t}$ becomes nongaussian and 
a first and second order description is not sufficient. 
Second, the noise will become nongaussian in time-discrete 
nonlinear models.
This results from the theory of integrating stochastic
differential equations, see \cite{kloeden91} for a recent review.
One main result of this theory is that unlike deterministic
differential equations, stochastic ones can not be integrated by
arbitrary high order methods, e.g.~by integrating the deterministic
part by a Runge-Kutta method of 4.th order and adding some noise
to the resulting value. Due to complicated stochastic integrals
in the Taylor expansion for higher order integration schemes, 
only first or second order methods are feasible. Therefore, usually,
the process has to be integrated on a much smaller time scale than 
it is observed. This turns the effective noise on the time scale
of the observation nongaussian. Because of this effect methods like
those proposed in \cite{borland92a,borland92b} are applicable in
special cases only.

The nongaussianity prevents
a maximum likelihood estimation of the parameter. A least square
approach yields biased estimates \cite{harvey90}. In general, for nongaussian
distributions all higher moments have to be taken into account.
A truncated expansion in moments was suggested in \cite{gelb89},
the application of the Gibbs sampler using a mixture of normals
to approximate the distributions was introduced in \cite{carlin92}.
To our knowledge no application of these methods to empirical
data have been reported.

All the map-based methods mentioned in Sec.~\ref{ndc}
can be applied
to nonlinear stochastic system by allowing for a residual error
in the prediction. Here, the above mentioned non-gaussianity is
also a problem since commonly applied least square fitting is
no longer maximum likelihood.

An application of the ''Error-in-variables''-theory
to this double stochastic situation is not straightforward.
To apply this theory, the ratio of the variance of the noise of past
values to that of the present value
must be known. The noise covering the past values is simply the observational
noise, but the noise on the present value is, in contrast to pure
regression, the sum of the observational and the dynamical noise. 
Since spectra of most (observational noise free) processes
 decay for high frequencies
\cite{sigeti87,sigeti95}, the variance of the observational noise can 
be estimated from this region. But the variance of the dynamical noise
is usually obtained after fitting the model since minimizing this
variance is the optimizing criterion.

The severity of the effects of the  problems discussed depends on the
 system, i.e.~the type of the nonlinearity and the variances of dynamical
and observational noise. 
We demonstrate this by briefly reporting the results of modeling a
time series of Parkinsonian
tremor. This type of tremor is believed to be a nonlinear
process \cite{gantert92,timmer93}. The time series was recorded under
the same conditions as the physiological tremor data analyzed in
Sec.~\ref{model_phys}. Fortunately, the observational noise in 
these data is negligible small.
Using a large variety of ansatzes for the right hand side of the 
differential equation, we did not succeed in fitting a nonlinear
deterministic differential equation by Bock's algorithm to the data.
Therefore, we assumed that the process is stochastic and tried
 global polynomials that are orthogonal with respect to
the measure given by the data. These polynomials were discussed in the context
of regression theory by Forsythe \cite{forsythe57}.
He also introduced a recursive algorithm to determine the polynomials
and their coefficients.
In \cite{giona91,brown93,aguirre95} they were made popular for the analysis
of time series. As discussed in \cite{aguirre95} the significance of
each single polynomial can be judged independently from the other
polynomials.

\begin{figure}
\begin{center}
\epsfxsize=0.6\textwidth 
\parbox{\textwidth}{
\epsfbox{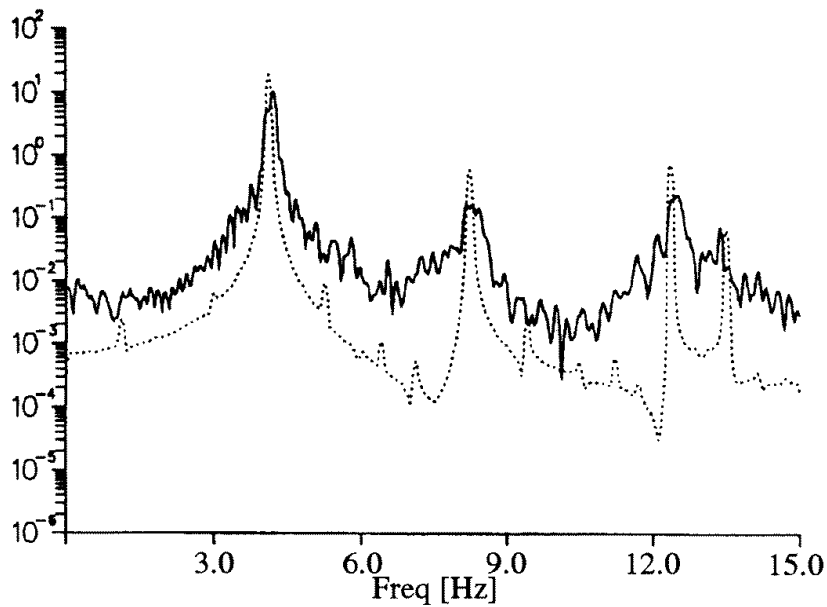}}\\ $\quad$ \\
\caption{\label{park_spec_fig}Spectrum estimated from the original
	data (solid line) and from a realization of the fitted
	model (dotted line).}
\end{center}
\end{figure}

Using  models of polynomial order 4 and an embedding delay $\tau$ of 
10 time steps we fitted models of ascending autoregression order to the data.
We found a model of regression order 3 containing 16 significant
parameters which describes the data adequately.
The goodness of fit was judged by a clear cut knee in the 
residual variance (Fig.~\ref{park_var_fig}), 
%
%------------------------------------------------------------
%  Fig.7 (eopfehl.tex) around here
%------------------------------------------------------------
%

\begin{center}
\begin{figure}
  \input {eopfehl.tex  }
    \caption{\label {park_var_fig}Variance of the prediction errors for
	4.th order polynomial models of ascending autoregression order.}
\end{figure}
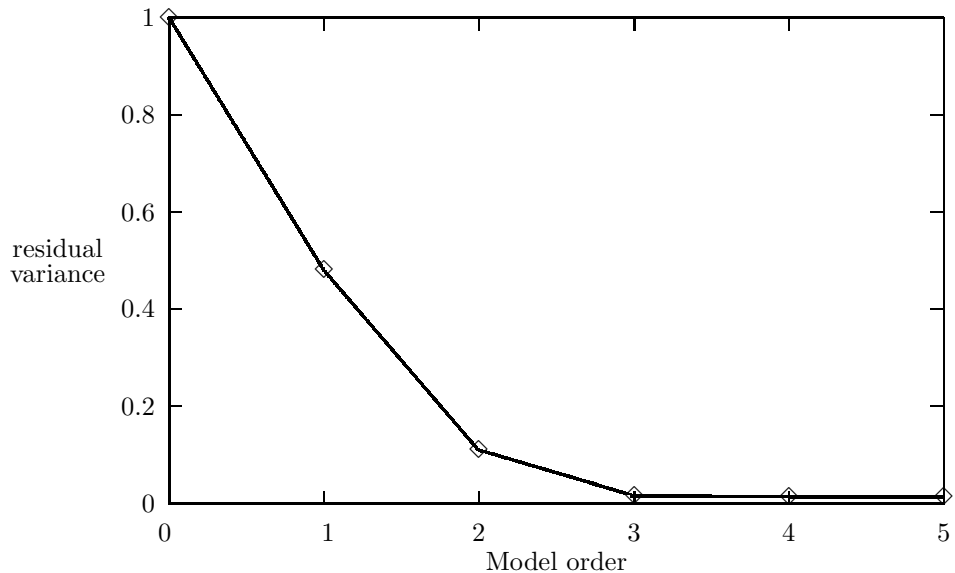
\end{center}

the whiteness of the residuals (Fig.~\ref{park_kolmo_fig}) and a comparison
of the spectra of the empirical data and data realized from 
%
%\marginpar {Please Fig.~\ref{park_kolmo_fig} around here}
%------------------------------------------------------------
%  Fig.8 (eopprob.tex) around here
%------------------------------------------------------------
%
\begin{center}
\begin{figure}
  \input {eopprob.tex  }
    \caption{\label{park_kolmo_fig} Kolmogorov-Smirnov test statistic D
	for the consistency of the prediction errors with white noise
	of 4.th order polynomials models of ascending autoregression
	order. The 5 \% significance level is marked.}
\end{figure}
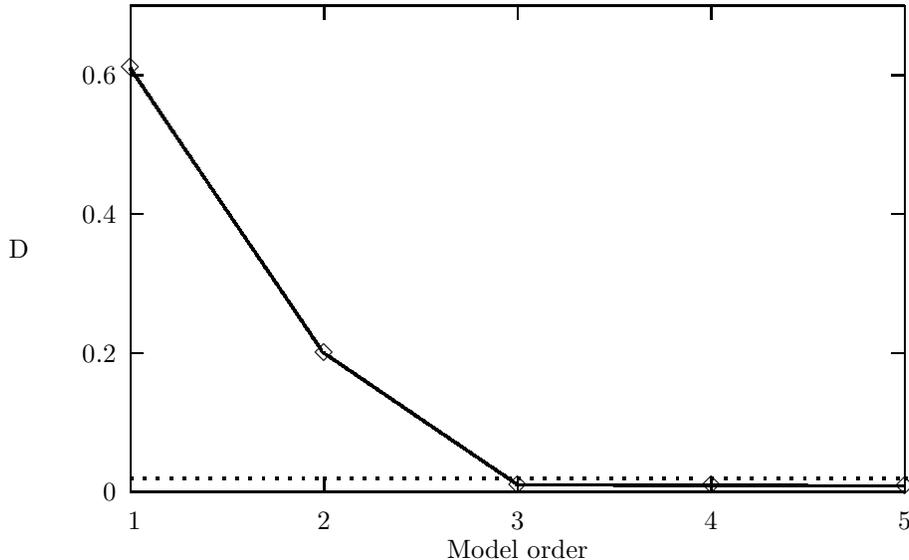
\end{center}

the model (Fig.~\ref{park_spec_fig}). Note, that in
Fig.~\ref{park_spec_fig} also the small peak near 13.5 Hz which is
only due to aliasing because of the effective downsampling by 
choosing $\tau= 10$, is reproduced by the model.

%
%\marginpar {Please Fig.~\ref{park_spec_fig} around here}
%------------------------------------------------------------
%  Fig.9 (spectra.eps) around here
%------------------------------------------------------------
%

Unfortunately we did not succeed in assigning a physiologic meaning
to the fitted parameters.

\section {  Discussion  }
The project ''Equations of Motion from a Data Series'' \cite{crutchfield87} 
has attracted much attention. Our own interest was to obtain models for 
physiological data like EEG or tremor. The hope was to learn something
about the underlying systems from the fitted models.
In the first part of this paper we gave an example
where this idea could be fulfilled and the
parameter of the fitted model could be interpreted in terms
of physics and physiology.
This was facilitated by the fact that the well understood theory of
linear systems was applicable.

In the general case of nonlinear deterministic or even nonlinear
stochastic systems it appears
to be much harder to obtain such a result. This is on the one hand
caused by the fact that the treatment of observational noise 
is not solved in general. On the other hand without much knowledge
about the system under investigation, it is hard to decide on 
a certain interpretable ansatz for the parameterization of the
nonlinear dynamics.

Often an interpretable model is not the goal of modeling, e.g.~if 
only prediction as for stock market data or the temporal
evolution of the prediction error as for discriminating 
deterministic from stochastic behavior is the aim.
But here also observational noise is a problem,
since the functional relationship will be underestimated
if it is modeled by maps. A model taking
this noise into account would lead to a better model and
a smaller prediction error.

Results for fitted differential equations might be compared 
to known systems in order to understand the underlying mechanisms. 
For maps this seems to be much more difficult, especially
when sigmoidals, radial basis functions or local linear models 
are used as basis
functions. But even if global polynomials are used an assignment
of a physical meaning to the parameters is difficult since a single
parameter in the underlying differential equation shows up
in the coefficients of more than one basis function. Furthermore,
the model structure depends on the sampling time.

For many methods applied to model time series, observational noise
causes problems since these methods treat the data as in
 regression. For deterministic systems Bock's algorithm
provides an elegant alternative which explores the additional
information that the process under investigation produced
a continuous trajectory.

There is growing evidence that many physiological processes
are neither linear stochastic processes since surrogate data testing 
reject this
hypothesis nor nonlinear deterministic processes since 
low dimensional attractors can not be established. Therefore,
we believe that methods to model (noisy) nonlinear stochastic processes are
worth being studied in more detail.

\section* {Data availability}

The tremor data and the simulated data of the Lotka-Volterra system
 are accessible at:\\
http://phym1.physik.uni-freiburg.de/$\sim$jeti/tremordata/

\section* {Acknowledgments}
We would like to thank M.~Lauk, T.~M\"uller
and A.~Weigend for valuable discussion on various aspects
of this paper. The tremor data were recorded at the
Neurology Department of the University Hospital Freiburg and were
kindly made available to us by C.H.~L\"ucking and G.~Deuschl. 
Special thanks to T.~M\"uller for implementing
Bock's algorithm and performing the simulations.

\bibliography{/u/honer/jeti/tex/paper/jetilit}

\bibliographystyle{plain}

\end {document}

%% file: FIG3.tex
% GNUPLOT: LaTeX picture
\setlength{\unitlength}{0.240900pt}
\ifx\plotpoint\undefined\newsavebox{\plotpoint}\fi
\begin{picture}(1500,900)(0,0)
\font\gnuplot=cmr10 at 10pt
\gnuplot
\sbox{\plotpoint}{\rule[-0.200pt]{0.400pt}{0.400pt}}%
\put(220.0,113.0){\rule[-0.200pt]{4.818pt}{0.400pt}}
\put(198,113){\makebox(0,0)[r]{0.6}}
\put(1416.0,113.0){\rule[-0.200pt]{4.818pt}{0.400pt}}
\put(220.0,304.0){\rule[-0.200pt]{4.818pt}{0.400pt}}
\put(198,304){\makebox(0,0)[r]{0.7}}
\put(1416.0,304.0){\rule[-0.200pt]{4.818pt}{0.400pt}}
\put(220.0,495.0){\rule[-0.200pt]{4.818pt}{0.400pt}}
\put(198,495){\makebox(0,0)[r]{0.8}}
\put(1416.0,495.0){\rule[-0.200pt]{4.818pt}{0.400pt}}
\put(220.0,686.0){\rule[-0.200pt]{4.818pt}{0.400pt}}
\put(198,686){\makebox(0,0)[r]{0.9}}
\put(1416.0,686.0){\rule[-0.200pt]{4.818pt}{0.400pt}}
\put(220.0,877.0){\rule[-0.200pt]{4.818pt}{0.400pt}}
\put(198,877){\makebox(0,0)[r]{1}}
\put(1416.0,877.0){\rule[-0.200pt]{4.818pt}{0.400pt}}
\put(220.0,113.0){\rule[-0.200pt]{0.400pt}{4.818pt}}
\put(220,68){\makebox(0,0){0}}
\put(220.0,857.0){\rule[-0.200pt]{0.400pt}{4.818pt}}
\put(463.0,113.0){\rule[-0.200pt]{0.400pt}{4.818pt}}
\put(463,68){\makebox(0,0){2}}
\put(463.0,857.0){\rule[-0.200pt]{0.400pt}{4.818pt}}
\put(706.0,113.0){\rule[-0.200pt]{0.400pt}{4.818pt}}
\put(706,68){\makebox(0,0){4}}
\put(706.0,857.0){\rule[-0.200pt]{0.400pt}{4.818pt}}
\put(950.0,113.0){\rule[-0.200pt]{0.400pt}{4.818pt}}
\put(950,68){\makebox(0,0){6}}
\put(950.0,857.0){\rule[-0.200pt]{0.400pt}{4.818pt}}
\put(1193.0,113.0){\rule[-0.200pt]{0.400pt}{4.818pt}}
\put(1193,68){\makebox(0,0){8}}
\put(1193.0,857.0){\rule[-0.200pt]{0.400pt}{4.818pt}}
\put(1436.0,113.0){\rule[-0.200pt]{0.400pt}{4.818pt}}
\put(1436,68){\makebox(0,0){10}}
\put(1436.0,857.0){\rule[-0.200pt]{0.400pt}{4.818pt}}
\put(220.0,113.0){\rule[-0.200pt]{292.934pt}{0.400pt}}
\put(1436.0,113.0){\rule[-0.200pt]{0.400pt}{184.048pt}}
\put(220.0,877.0){\rule[-0.200pt]{292.934pt}{0.400pt}}
\put(45,495){\makebox(0,0){\shortstack{\shortstack{Residual\\variance}}}}
\put(828,23){\makebox(0,0){Model order}}
\put(220.0,113.0){\rule[-0.200pt]{0.400pt}{184.048pt}}
\put(220,877){\usebox{\plotpoint}}
\multiput(220.58,869.25)(0.499,-2.214){241}{\rule{0.120pt}{1.867pt}}
\multiput(219.17,873.12)(122.000,-535.125){2}{\rule{0.400pt}{0.934pt}}
\multiput(342.58,335.65)(0.499,-0.583){239}{\rule{0.120pt}{0.566pt}}
\multiput(341.17,336.83)(121.000,-139.825){2}{\rule{0.400pt}{0.283pt}}
\put(828,195.67){\rule{29.390pt}{0.400pt}}
\multiput(828.00,196.17)(61.000,-1.000){2}{\rule{14.695pt}{0.400pt}}
\put(463.0,197.0){\rule[-0.200pt]{87.928pt}{0.400pt}}
\put(1071,194.67){\rule{29.390pt}{0.400pt}}
\multiput(1071.00,195.17)(61.000,-1.000){2}{\rule{14.695pt}{0.400pt}}
\put(950.0,196.0){\rule[-0.200pt]{29.149pt}{0.400pt}}
\put(1193.0,195.0){\rule[-0.200pt]{58.539pt}{0.400pt}}
\put(220,877){\raisebox{-.8pt}{\makebox(0,0){$\Diamond$}}}
\put(342,338){\raisebox{-.8pt}{\makebox(0,0){$\Diamond$}}}
\put(463,197){\raisebox{-.8pt}{\makebox(0,0){$\Diamond$}}}
\put(585,197){\raisebox{-.8pt}{\makebox(0,0){$\Diamond$}}}
\put(706,197){\raisebox{-.8pt}{\makebox(0,0){$\Diamond$}}}
\put(828,197){\raisebox{-.8pt}{\makebox(0,0){$\Diamond$}}}
\put(950,196){\raisebox{-.8pt}{\makebox(0,0){$\Diamond$}}}
\put(1071,196){\raisebox{-.8pt}{\makebox(0,0){$\Diamond$}}}
\put(1193,195){\raisebox{-.8pt}{\makebox(0,0){$\Diamond$}}}
\put(1314,195){\raisebox{-.8pt}{\makebox(0,0){$\Diamond$}}}
\put(1436,195){\raisebox{-.8pt}{\makebox(0,0){$\Diamond$}}}
\put(220,877){\usebox{\plotpoint}}
\multiput(220,877)(6.406,-19.742){20}{\usebox{\plotpoint}}
\multiput(342,501)(15.107,-14.233){8}{\usebox{\plotpoint}}
\multiput(463,387)(19.421,-7.323){6}{\usebox{\plotpoint}}
\multiput(585,341)(20.504,-3.220){6}{\usebox{\plotpoint}}
\multiput(706,322)(20.749,-0.510){6}{\usebox{\plotpoint}}
\multiput(828,319)(20.756,0.000){5}{\usebox{\plotpoint}}
\multiput(950,319)(20.744,-0.686){6}{\usebox{\plotpoint}}
\multiput(1071,315)(20.699,-1.527){6}{\usebox{\plotpoint}}
\multiput(1193,306)(20.670,-1.879){6}{\usebox{\plotpoint}}
\multiput(1314,295)(20.656,-2.032){6}{\usebox{\plotpoint}}
\put(1436,283){\usebox{\plotpoint}}
\put(220,877){\makebox(0,0){$+$}}
\put(342,501){\makebox(0,0){$+$}}
\put(463,387){\makebox(0,0){$+$}}
\put(585,341){\makebox(0,0){$+$}}
\put(706,322){\makebox(0,0){$+$}}
\put(828,319){\makebox(0,0){$+$}}
\put(950,319){\makebox(0,0){$+$}}
\put(1071,315){\makebox(0,0){$+$}}
\put(1193,306){\makebox(0,0){$+$}}
\put(1314,295){\makebox(0,0){$+$}}
\put(1436,283){\makebox(0,0){$+$}}
\end{picture}

%% file: eopfehl.tex
% GNUPLOT: LaTeX picture
\setlength{\unitlength}{0.240900pt}
\ifx\plotpoint\undefined\newsavebox{\plotpoint}\fi
\begin{picture}(1500,900)(0,0)
\font\gnuplot=cmr10 at 10pt
\gnuplot
\sbox{\plotpoint}{\rule[-0.200pt]{0.400pt}{0.400pt}}%
\put(220.0,113.0){\rule[-0.200pt]{4.818pt}{0.400pt}}
\put(198,113){\makebox(0,0)[r]{0}}
\put(1416.0,113.0){\rule[-0.200pt]{4.818pt}{0.400pt}}
\put(220.0,266.0){\rule[-0.200pt]{4.818pt}{0.400pt}}
\put(198,266){\makebox(0,0)[r]{0.2}}
\put(1416.0,266.0){\rule[-0.200pt]{4.818pt}{0.400pt}}
\put(220.0,419.0){\rule[-0.200pt]{4.818pt}{0.400pt}}
\put(198,419){\makebox(0,0)[r]{0.4}}
\put(1416.0,419.0){\rule[-0.200pt]{4.818pt}{0.400pt}}
\put(220.0,571.0){\rule[-0.200pt]{4.818pt}{0.400pt}}
\put(198,571){\makebox(0,0)[r]{0.6}}
\put(1416.0,571.0){\rule[-0.200pt]{4.818pt}{0.400pt}}
\put(220.0,724.0){\rule[-0.200pt]{4.818pt}{0.400pt}}
\put(198,724){\makebox(0,0)[r]{0.8}}
\put(1416.0,724.0){\rule[-0.200pt]{4.818pt}{0.400pt}}
\put(220.0,877.0){\rule[-0.200pt]{4.818pt}{0.400pt}}
\put(198,877){\makebox(0,0)[r]{1}}
\put(1416.0,877.0){\rule[-0.200pt]{4.818pt}{0.400pt}}
\put(220.0,113.0){\rule[-0.200pt]{0.400pt}{4.818pt}}
\put(220,68){\makebox(0,0){0 }}
\put(220.0,857.0){\rule[-0.200pt]{0.400pt}{4.818pt}}
\put(463.0,113.0){\rule[-0.200pt]{0.400pt}{4.818pt}}
\put(463,68){\makebox(0,0){ 1}}
\put(463.0,857.0){\rule[-0.200pt]{0.400pt}{4.818pt}}
\put(706.0,113.0){\rule[-0.200pt]{0.400pt}{4.818pt}}
\put(706,68){\makebox(0,0){2}}
\put(706.0,857.0){\rule[-0.200pt]{0.400pt}{4.818pt}}
\put(950.0,113.0){\rule[-0.200pt]{0.400pt}{4.818pt}}
\put(950,68){\makebox(0,0){3}}
\put(950.0,857.0){\rule[-0.200pt]{0.400pt}{4.818pt}}
\put(1193.0,113.0){\rule[-0.200pt]{0.400pt}{4.818pt}}
\put(1193,68){\makebox(0,0){4}}
\put(1193.0,857.0){\rule[-0.200pt]{0.400pt}{4.818pt}}
\put(1436.0,113.0){\rule[-0.200pt]{0.400pt}{4.818pt}}
\put(1436,68){\makebox(0,0){5}}
\put(1436.0,857.0){\rule[-0.200pt]{0.400pt}{4.818pt}}
\put(220.0,113.0){\rule[-0.200pt]{292.934pt}{0.400pt}}
\put(1436.0,113.0){\rule[-0.200pt]{0.400pt}{184.048pt}}
\put(220.0,877.0){\rule[-0.200pt]{292.934pt}{0.400pt}}
\put(45,495){\makebox(0,0){\shortstack{\shortstack{residual\\variance}}}}
\put(828,23){\makebox(0,0){Model order}}
\put(220.0,113.0){\rule[-0.200pt]{0.400pt}{184.048pt}}
\sbox{\plotpoint}{\rule[-0.400pt]{0.800pt}{0.800pt}}%
\put(220,877){\usebox{\plotpoint}}
\multiput(221.41,870.74)(0.500,-0.817){479}{\rule{0.121pt}{1.507pt}}
\multiput(218.34,873.87)(243.000,-393.872){2}{\rule{0.800pt}{0.753pt}}
\multiput(464.41,475.30)(0.500,-0.582){479}{\rule{0.121pt}{1.132pt}}
\multiput(461.34,477.65)(243.000,-280.651){2}{\rule{0.800pt}{0.566pt}}
\multiput(706.00,195.09)(1.705,-0.501){137}{\rule{2.911pt}{0.121pt}}
\multiput(706.00,195.34)(237.958,-72.000){2}{\rule{1.456pt}{0.800pt}}
\put(950,122.84){\rule{58.539pt}{0.800pt}}
\multiput(950.00,123.34)(121.500,-1.000){2}{\rule{29.269pt}{0.800pt}}
\put(1193,121.84){\rule{58.539pt}{0.800pt}}
\multiput(1193.00,122.34)(121.500,-1.000){2}{\rule{29.269pt}{0.800pt}}
\sbox{\plotpoint}{\rule[-0.200pt]{0.400pt}{0.400pt}}%
\put(220,877){\raisebox{-.8pt}{\makebox(0,0){$\Diamond$}}}
\put(463,480){\raisebox{-.8pt}{\makebox(0,0){$\Diamond$}}}
\put(706,197){\raisebox{-.8pt}{\makebox(0,0){$\Diamond$}}}
\put(950,125){\raisebox{-.8pt}{\makebox(0,0){$\Diamond$}}}
\put(1193,124){\raisebox{-.8pt}{\makebox(0,0){$\Diamond$}}}
\put(1436,123){\raisebox{-.8pt}{\makebox(0,0){$\Diamond$}}}
\end{picture}

%% file: eopprob.tex
% GNUPLOT: LaTeX picture
\setlength{\unitlength}{0.240900pt}
\ifx\plotpoint\undefined\newsavebox{\plotpoint}\fi
\begin{picture}(1500,900)(0,0)
\font\gnuplot=cmr10 at 10pt
\gnuplot
\sbox{\plotpoint}{\rule[-0.200pt]{0.400pt}{0.400pt}}%
\put(220.0,113.0){\rule[-0.200pt]{4.818pt}{0.400pt}}
\put(198,113){\makebox(0,0)[r]{0}}
\put(1416.0,113.0){\rule[-0.200pt]{4.818pt}{0.400pt}}
\put(220.0,331.0){\rule[-0.200pt]{4.818pt}{0.400pt}}
\put(198,331){\makebox(0,0)[r]{0.2}}
\put(1416.0,331.0){\rule[-0.200pt]{4.818pt}{0.400pt}}
\put(220.0,550.0){\rule[-0.200pt]{4.818pt}{0.400pt}}
\put(198,550){\makebox(0,0)[r]{0.4}}
\put(1416.0,550.0){\rule[-0.200pt]{4.818pt}{0.400pt}}
\put(220.0,768.0){\rule[-0.200pt]{4.818pt}{0.400pt}}
\put(198,768){\makebox(0,0)[r]{0.6}}
\put(1416.0,768.0){\rule[-0.200pt]{4.818pt}{0.400pt}}
\put(220.0,113.0){\rule[-0.200pt]{0.400pt}{4.818pt}}
\put(220,68){\makebox(0,0){ 1}}
\put(220.0,857.0){\rule[-0.200pt]{0.400pt}{4.818pt}}
\put(524.0,113.0){\rule[-0.200pt]{0.400pt}{4.818pt}}
\put(524,68){\makebox(0,0){2}}
\put(524.0,857.0){\rule[-0.200pt]{0.400pt}{4.818pt}}
\put(828.0,113.0){\rule[-0.200pt]{0.400pt}{4.818pt}}
\put(828,68){\makebox(0,0){3}}
\put(828.0,857.0){\rule[-0.200pt]{0.400pt}{4.818pt}}
\put(1132.0,113.0){\rule[-0.200pt]{0.400pt}{4.818pt}}
\put(1132,68){\makebox(0,0){4}}
\put(1132.0,857.0){\rule[-0.200pt]{0.400pt}{4.818pt}}
\put(1436.0,113.0){\rule[-0.200pt]{0.400pt}{4.818pt}}
\put(1436,68){\makebox(0,0){5}}
\put(1436.0,857.0){\rule[-0.200pt]{0.400pt}{4.818pt}}
\put(220.0,113.0){\rule[-0.200pt]{292.934pt}{0.400pt}}
\put(1436.0,113.0){\rule[-0.200pt]{0.400pt}{184.048pt}}
\put(220.0,877.0){\rule[-0.200pt]{292.934pt}{0.400pt}}
\put(45,495){\makebox(0,0){D}}
\put(828,23){\makebox(0,0){Model order}}
\put(220.0,113.0){\rule[-0.200pt]{0.400pt}{184.048pt}}
\sbox{\plotpoint}{\rule[-0.400pt]{0.800pt}{0.800pt}}%
\put(220,779){\usebox{\plotpoint}}
\multiput(221.41,773.28)(0.500,-0.737){601}{\rule{0.121pt}{1.379pt}}
\multiput(218.34,776.14)(304.000,-445.138){2}{\rule{0.800pt}{0.689pt}}
\multiput(524.00,329.09)(0.735,-0.500){407}{\rule{1.375pt}{0.121pt}}
\multiput(524.00,329.34)(301.146,-207.000){2}{\rule{0.687pt}{0.800pt}}
\put(828,121.84){\rule{73.234pt}{0.800pt}}
\multiput(828.00,122.34)(152.000,-1.000){2}{\rule{36.617pt}{0.800pt}}
\put(1132,120.84){\rule{73.234pt}{0.800pt}}
\multiput(1132.00,121.34)(152.000,-1.000){2}{\rule{36.617pt}{0.800pt}}
\sbox{\plotpoint}{\rule[-0.200pt]{0.400pt}{0.400pt}}%
\put(220,779){\raisebox{-.8pt}{\makebox(0,0){$\Diamond$}}}
\put(524,331){\raisebox{-.8pt}{\makebox(0,0){$\Diamond$}}}
\put(828,124){\raisebox{-.8pt}{\makebox(0,0){$\Diamond$}}}
\put(1132,123){\raisebox{-.8pt}{\makebox(0,0){$\Diamond$}}}
\put(1436,122){\raisebox{-.8pt}{\makebox(0,0){$\Diamond$}}}
\sbox{\plotpoint}{\rule[-0.500pt]{1.000pt}{1.000pt}}%
\put(220,134){\usebox{\plotpoint}}
\multiput(220,134)(20.756,0.000){15}{\usebox{\plotpoint}}
\multiput(524,134)(20.756,0.000){15}{\usebox{\plotpoint}}
\multiput(828,134)(20.756,0.000){14}{\usebox{\plotpoint}}
\multiput(1132,134)(20.756,0.000){15}{\usebox{\plotpoint}}
\put(1436,134){\usebox{\plotpoint}}
\end{picture}